\documentclass[useAMS,usenatbib,twoside,a4paper]{mn2e}
\usepackage{graphicx}
\title[New X-ray pulsars in the SMC]
{Three new X-ray pulsars detected in the Small Magellanic Cloud and the positions of two other known
pulsars determined.}
\author[W. R. T. Edge et al.]
       {W. R. T. Edge,$^1$ M. J. Coe,$^1$ J. L. Galache,$^1$ V. A. McBride,$^1$
       \newauthor R.H.D. Corbet,$^2{}^,{}^3$ C.B. Markwardt,$^2{}^,{}^4$ S.
Laycock$^5$
\\
        $^1$School of Physics and Astronomy, Southampton University, SO17
        1BJ,\\
        $^2$NASA Goddard Space Flight Center, Greenbelt, MD 20771 USA\\
        $^3$Universities Space Research Association\\
        $^4$Department of Astronomy, University of Maryland, College Park, MD 20742, USA\\
        $^5$Harvard-Smithsonian Center for Astrophysics, Cambridge, MA 02138, USA}

\begin{document}

\date{Accepted 2004.
      Received 2004;
      in original form 2004}

\pagerange{\pageref{firstpage}--\pageref{lastpage}} \pubyear{2003}

\maketitle

\label{firstpage}

\begin{abstract}

Three new X-ray pulsars have been detected in the Small Magellanic
Cloud (SMC) and the positions of two others have been determined,
with archive Chandra data. A series of five observations of the
SMC took place between May and October 2002. Analysis of these
data has revealed three previously unknown X-ray pulsars at pulse
periods of 34, 503 and 138 seconds. The position of pulsar XTE
J0052-725, which was originally detected by RXTE on June 19 2002,
was also accurately determined and a previously detected 7.78s
RXTE pulsar was identified as the source SMC X-3.

\end{abstract}


\section{INTRODUCTION}

The Magellanic Clouds are a pair of satellite galaxies which are gravitationally bound to our own but
which have structural and chemical characteristics differing significantly from each other, and from
the Milky Way. These differences are likely to be reflected in the properties of different stellar
populations. The Small Magellanic cloud (SMC) is located at a distance of about 60 kpc (Harries,
Hilditch, \& Howarth, 2003) and centred on a position of R.A. 1hr Dec. -73$^{o}$. It is therefore
close enough to be observed with modest ground based telescopes whilst at the same time providing an
opportunity to study and compare the evolution of other galaxies.

Intensive X-ray satellite observations have revealed that the SMC
contains an unexpectedly large number of High Mass X-ray Binaries
(HMXB). At the time of writing, 46 known or probable sources of
this type have been identified in the SMC and they continue to be
discovered at a rate of about 1-3 per year, although only a small
fraction of these are active at any one time because of their
transient nature. All X-ray binaries so far discovered in the SMC
are HMXBs (Coe 2000).

Most High Mass X-ray Binaries (HMXBs) belong to the Be class, in which a neutron star orbits an OB
star surrounded by a circumstellar disk of variable size and density. The optical companion stars are
early-type O-B class stars of luminosity class III-V, typically of 10 to 20 solar masses that at some
time have shown emission in the Balmer series lines. The systems as a whole exhibit significant
excess flux at long (IR and radio) wavelengths, referred to as the infrared excess. These
characteristic signatures as well as strong H$\alpha $ line emission are attributed to the presence
of circumstellar material in a disk-like configuration (Coe 2000, Okazaki and Negueruela 2001).

The mechanisms which give rise to the disk are not well understood, although fast rotation is likely
to be an important factor, and it is possible that non-radial pulsation and magnetic loops may also
play a part. Short-term periodic variability is observed in the earlier type Be stars.The disk is
thought to consist of relatively cool material, which interacts periodically with a compact object in
an eccentric orbit, leading to regular X-ray outbursts. It is also possible that the Be star
undergoes a sudden ejection of matter (Negueruela 1998, Porter \& Rivinius, 2003).

Be/X-ray binaries can present differing states of X-ray activity
varying from persistent low or non-detectable luminosities to
short outbursts. Systems with wide orbits will tend to accrete
from less dense regions of the disk and hence show relatively
small outbursts. These are referred to as Type I and usually
coincide with the periastron of the neutron star. Systems with
smaller orbits are more likely to accrete from dense regions over
a range of orbital phases and give rise to very high luminosity
outbursts, although these may be modulated by the presence of a
density wave in the disk. Prolonged major outbursts, which do not
exhibit signs of orbital modulation, are normally called Type II
(Negueruela 1998).

\section{The Chandra Data}

\begin{figure}\begin{center}
\includegraphics[width=85mm]{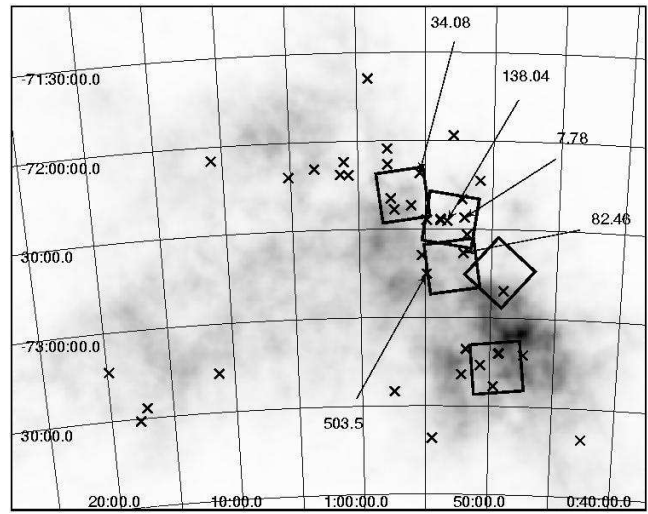}
\caption{Positions covered by the ACIS-I arrays from Chandra observations 2944 - 2948, overlaid on a
neutral hydrogen density image of the SMC. The newly identified pulsars are indicated by their pulse
period, those already known are shown with a cross.} \label{fig:smc}
\end{center}
\end{figure}

A Chandra survey of the SMC bar region, initiated by Zezas et al., covered five separate fields
between May and October 2002 (Zezas et al., 2003). The positions covered by the ACIS-I arrays from
these observations are shown in relation to known pulsars in Figure~\ref{fig:smc}. These have been
overlaid on a neutral hydrogen density image of the SMC (Stanimirovi\'{c} et al. 1999).

The standard ACIS-I0123S23 CCD configuration was used although all sources discussed in this paper
were found in the ACIS-I region which provides a $16.9'\times16.9'$ field of view. Exposure times
were between 7.6 and 11.6~ksec. Data were extracted from the ACIS level 2 event fits files which were
taken with a frame readout time of 3.241 s. Using the CIAO v3 software, these observations were
barycentrically corrected after which potential sources were detected using the \textit{wavdetect}
algorithm. Background was subtracted using rectangular regions of varying sizes, after which timing
analysis of the lightcurves was carried out using both Lomb-Scargle and Fourier Transform algorithms.
The Starlink PERIOD programme was used to generate pulse profiles folded on the resultant periods and
these were then fitted with a sine function from which the pulsed fractions were determined.

Three of these observations (2946, 2947 and 2948) yielded significant results in the detection and
identification of X-ray pulsars. These are summarised in Table~\ref{tab1} which also includes two
previously known sources seen in the same data. The positional errors generated by \textit{wavdetect}
have been combined in quadrature with the Chandra nominal 90\% confidence radial uncertainty of 0.6
arcsec to give the errors shown.

Spectral information for these sources was obtained using the CIAO \textit{psextract} tool. The
spectra were initially binned to a minimum of 10 counts per bin using the FTOOLS utility GRPPHA
before being analysed with XSPEC. It was found however, that in the case of the three weakest sources
this resulted in a poor fit and these were therefore binned to 2 counts per bin. Channel energy
outside the limits of 0.3 keV and 10 keV was ignored. The spectra were then fitted to an absorbed
powerlaw model, the results of which are summarised in Table~\ref{tab2}. $\chi^{2}_{v}$ for the
weaker sources would not be statistically valid, because of the relatively low number of counts, and
has been omitted. The spectra are shown in Figure~\ref{fig:spectra}.

\begin{table*}
\begin{minipage}{150mm}
\begin{tabular}{|c|c|c|c|c|c|c|c|c|c|}
\hline Obj.&Name& R.A.& Dec.&error& P$_{pulse}$&Counts&Pulsed fract (\%) & ObsID&Date \\
  &       &   \textit{(J2000)}    &     &  \textit{(arcsecs)} &  \textit{(s)}  &     &       &      &       \\
\hline 1& XTE J0052-725 0.3-2.5 keV&00:52:08.9& -72:38:03&0.61& 82.46&3074&19$\pm$3 & 2946&04/Jul/02 \\
\hline 1& XTE J0052-725 2.5-10 KeV&    &     &   &   &2168&42$\pm$3 &   &     \\
\hline 2& SMC X-3&00:52:05.7& -72:26:04&0.62& 7.78&2103&27$\pm$3 & 2947&20/07/02 \\
\hline 3& CXOU J005455.6-724510& 00:54:55.8& -72:45:11&0.82&503.5&518&63$\pm$12&2946&04/Jul/02 \\
\hline 4& CXOU J005527.9-721058&00:55:27.7& -72:10:59&0.99& 34.08&293&57$\pm$23&2948&04/Jul/02 \\
\hline 5& CXOU J005323.8-722715&00:53:24.0& -72:27:16&1.00&138.04&166&59$\pm$25&2947&20/Jul/02 \\
\hline 6 & CXOU J005736.2-721934&00:57:35.9&-72:19:35&0.71&562&118&73$\pm$50&2948&04/Jul/02\\
\hline 7 & RX J0050.8-7316&00:50:44.6&-73:16:05&0.64&319.7&417&41$\pm$20&2945&02/Oct/02 \\
\hline
\end{tabular}
\caption{X-ray Binary sources detected in Chandra observations
2946, 2947 and 2948.} \label{tab1}
\end{minipage}
\end{table*}

\section{Individual Sources}

\subsection{XTE J0052-725}

\begin{figure*}
\begin{center} \hbox{
\includegraphics[height=22cm]{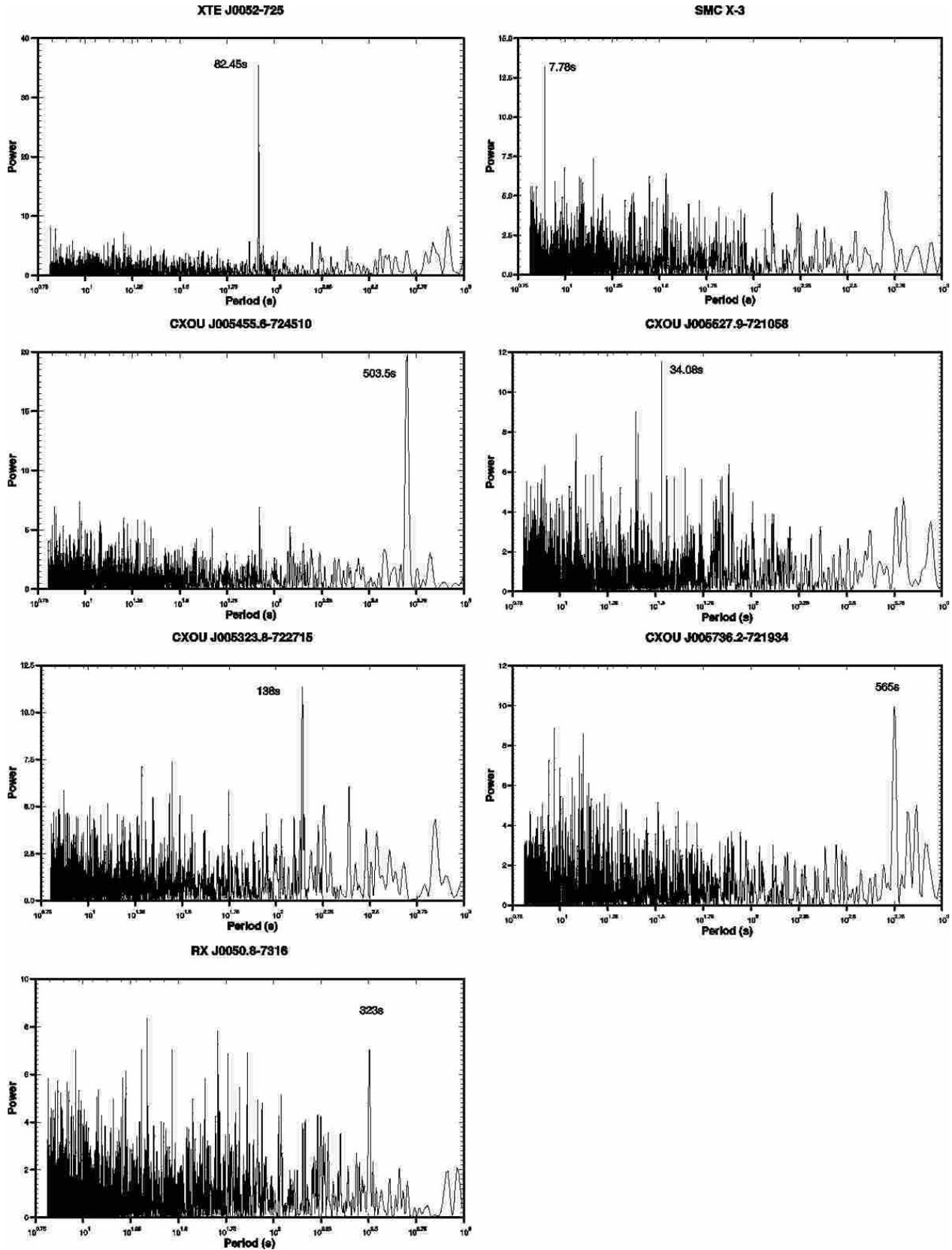}}
\end{center}
\caption{Power curves of all seven sources showing main periods detected using the Lomb-Scargle
algorithm in Starlink PERIOD.} \label{fig:powcurves}
\end{figure*}

Pulsar XTE J0052-725 was originally detected by RXTE on June 19 and 26, 2002 (Corbet et al., 2002).
The Chandra data show an X-ray source in Observation ID 2946 which took place on 4 July 2002 (MJD
52459) between 06:47:00 and 10:37:01 hrs. Timing analysis revealed a period of $82.46 \pm 0.18 s$ at
a confidence level of $>99$\%. This period is within 0.06s of the one detected by RXTE and the
position lies 6 arc minutes from the original mean RXTE location. It is therefore concluded that
these are the same pulsar. The Chandra power spectrum is shown in Figure~\ref{fig:powcurves}.

\begin{figure*}
\begin{center} \hbox{
\includegraphics[height=22cm]{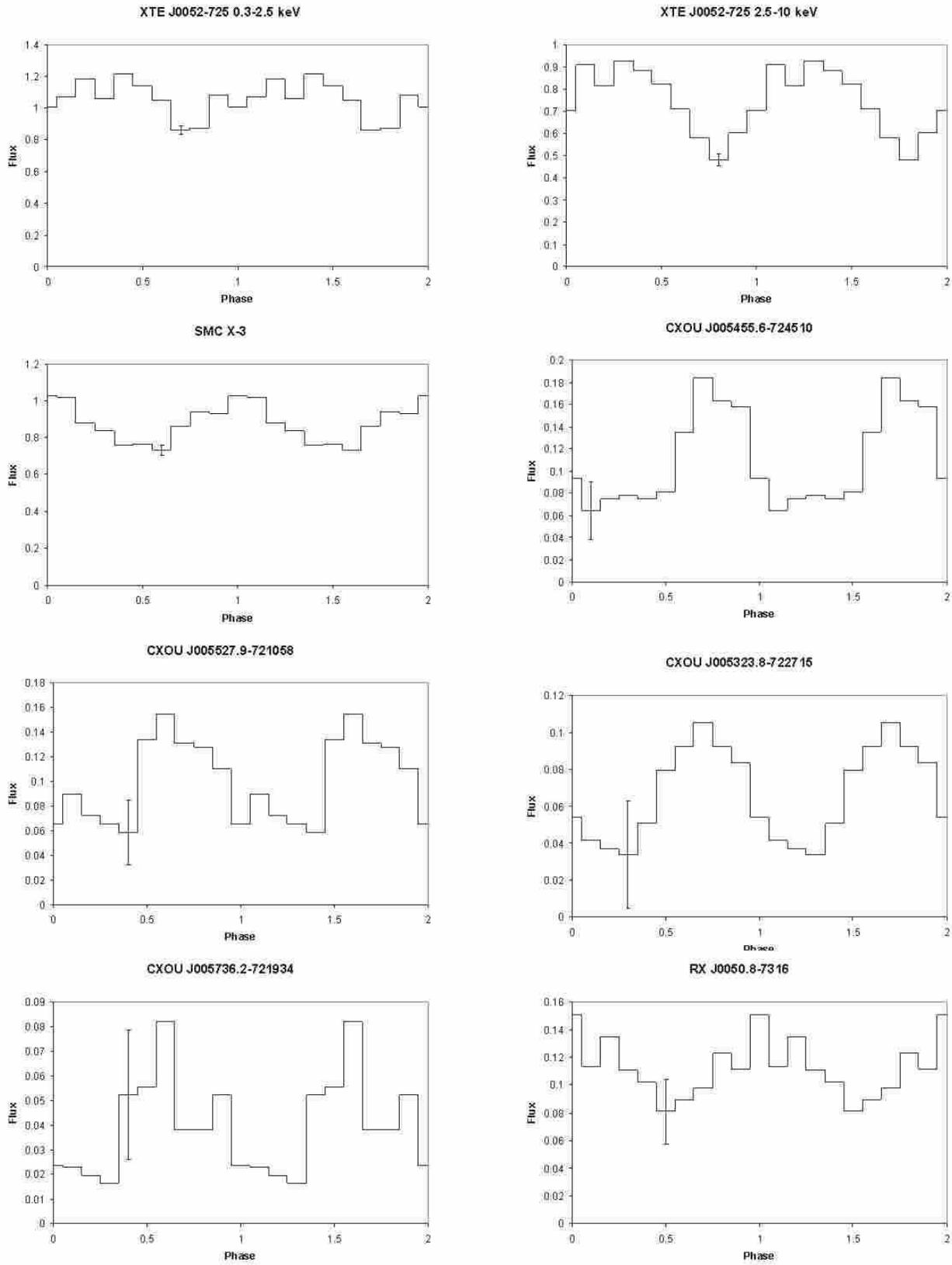}}
\end{center}
\caption{Pulse profiles of all seven sources with representative error bars.The top two figures show
the low and high energy curves of XTE J0052-725 separately.} \label{fig:profiles}
\end{figure*}

\begin{figure*}
\begin{center} \hbox{
\includegraphics[height=22cm]{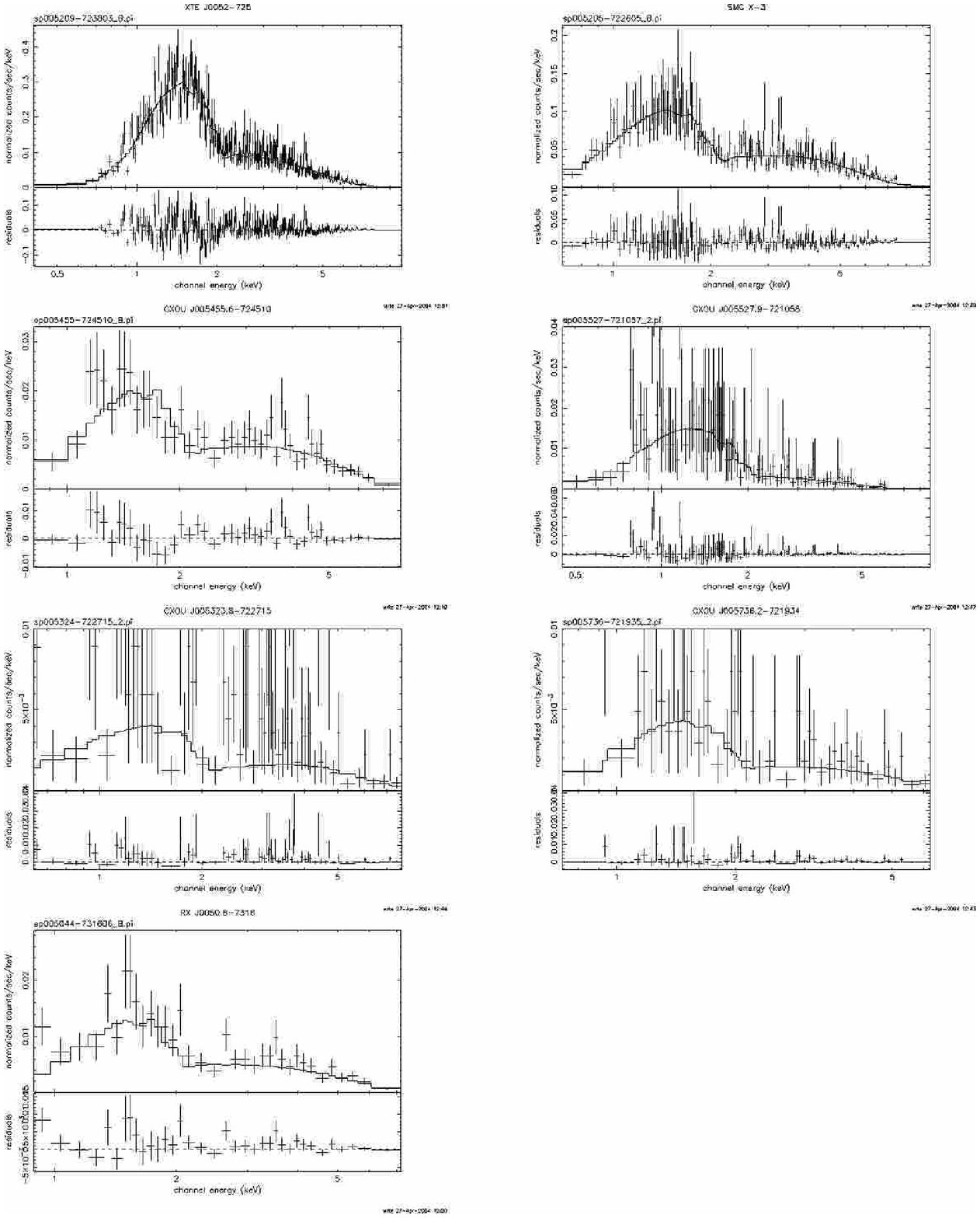}}
\end{center}
\caption{X-ray spectra of all seven sources. CXOU J005527.9-721058, CXOU J005323.8-722715 and CXOU
J005736.2-721934 have been binned to 2 counts per bin, the remainder to 10 counts per bin.}
\label{fig:spectra}
\end{figure*}

This source registered a total of 5255 counts and was therefore bright enough to test the pulse
profiles for energy dependance. Folded profiles for the 0.3-2.5 keV and 2.5-10 keV energy bands are
shown in Figure~\ref{fig:profiles}. The lower energy band contained about 60\% of the photons but had
a pulsed fraction of only $28$\% $\pm$ $2$\% as compared to $42$\% $\pm$ $3$\% in the higher energy
40\%. The spectrum, fitted to an absorbed powerlaw model, is in Figure~\ref{fig:spectra}. Assuming a
distance of 60 kpc to the SMC, the luminosity was 34.2 $\times$ 10$^{35}$ ergs s$^{-1}$. The
relatively high value of $N_{H}$ at 0.69 $\times$ $10^{22}$ $cm^{-2}$ may imply that this source is
at a greater depth into the SMC, and hence at a greater distance, or may simply reflect a greater
neutral hydrogen density in that region. This source has been identified with the optical counterpart
MACS J0052-726\#004 (Tucholke, de Boer, \& Seitter, 1996).

\subsection{SMC X-3}

SMC X-3, which was first detected in 1978(Clark et al. 1978), has now been identified from the
Chandra data with a previously detected 7.78s RXTE pulsar. Observation ID 2947, which took place on
20 July 2002 (MJD 52475) between 23:03:50 and 01:46:41 hrs, shows an X-ray source consistent with the
position of the optical counterpart proposed for SMC-X-3 by Crampton et al., 1978. Timing analysis
shows the Chandra object to have a pulse period of $7.781 \pm 0.002s$ with a confidence of $>98$\%.
An X-ray pulsar with a pulse period of $7.781 \pm 0.002s$ was detected by RXTE in early 2002 and on
eight subsequent occasions, giving a probable binary period of $45.1 \pm 0.4$ days (Corbet et al.,
2003). The position of the RXTE source was tentatively determined to be within 15 arcmins of the
known position of SMC X-3. An examination of the RXTE data shows that an outburst was detected in an
observation which took place on MJD 52478, three days after the Chandra observation on MJD 52475. It
is concluded that all these observations are of the same object, namely SMC X-3.

Analysis of the spectrum implies a luminosity of 27.6 $\times$ $10^{35}$ ergs s$^{-1}$ at 60 kpc. The
photon index of 0.7 indicates a relatively hard spectrum.

\subsection{CXOU J005455.6-724510}

An X-ray source in Observation ID 2946, which took place on 4 July 2002 (MJD 52459) between 06:48:04
and 09:47:57 was found to have a period of $503.5 \pm 6.7 s$ at a $>99$\% level of confidence.
Selective analysis of the source region was carried out to ensure that the period was not a harmonic
of the 1000s Y dithering frequency. The source was subsequently independently identified using XMM
data by Haberl et al. (2004) who computed the pulse period at $499.2 \pm 0.7s$. This object is very
close to RX J0054.9-7245 = AX J0054.8-7244 which is listed by both Haberl \& Pietsch (Haberl \&
Pietsch 2004) and Yokogawa et al. (Yokogawa et al., 2003) as a HMXB pulsar candidate.

\subsection{CXOU J005527.9-721058}

Observation ID 2948, took place on 4 July 2002 (MJD 52459) between 09:47:57 and 12:45:44. Timing
analysis on this object revealed a period of $34.08 \pm 0.03 s$ with a confidence of 98.5\%. The
position of this pulsar is within 3 arcsec of the ROSAT source 2RXP J005527.1-721100 (Rosat, 2000).
The latter is co-incident with a 16.8 V magnitude optical source having a B-V colour index of -0.116
(Zaritsky et al.2002) which would be consistent with the value expected from the optical companion in
a BeX-ray binary pair. To obtain the spectrum the data were first binned to 2 counts per bin.

\subsection{CXOU J005323.8-722715}

Observation ID 2947, took place on 20 July 2002 (MJD 52475) between 23:03:50 and 01:46:41. Timing
analysis on this object revealed a period of $138.04 \pm 0.61 s$ with a confidence of 98\%. The
position of this pulsar is coincident with emission-line star [MA93] 667 (Meyssonnier \& Azzopardi,
1993)and also with MACHO object 207.16202.50. The latter shows evidence of a period of $125 \pm 1.5$
days. This period would be consistent with that predicted from the Corbet diagram (Corbet 1986) for a
138s Be/X-ray pulsar.

The spectrum was obtained from data binned to 2 counts per bin. It is evident that the absorbed
powerlaw model gives a bad fit to the spectrum which may account for the apparent absence of $N_{H}$
(less than 0.01 $\times 10^{22}cm ^{-2}$). A broken powerlaw model was also tried and this gave a
better visual fit to the data points however the $N_{H}$ value disappeared to effectively nothing
which is clearly not consistent with a source located in the the SMC.

\begin{table*}
\begin{minipage}{120mm}
\begin{tabular}{|c|c|c|c|c|c|c|c|c|c|}
\hline Obj.  &    Name   &    L(Abs)X       &  $\chi ^{2}_{v}$ & $N_{dof}$ &$N_{H}$            & error &   PhoIndex  &error   \\
             &           &$10^{35} ergs/sec$&                  &           &$10^{22}cm ^{-2}$  &       &   ($\Gamma$)&        \\
\hline 1& XTE J0052-725& 34.19& 1.11& 269& 0.69& 0.03& 1.60&0.05 \\
\hline 2& SMC X-3& 27.58& 0.85& 159& 0.32& 0.05& 0.70&0.07 \\
\hline 3& CXOU J005455.6-724510& 5.54&0.80&43&0.56&0.15&0.84&0.16 \\
\hline 4& CXOU J005527.9-721058&1.10    &           &  &   0.30    &   0.09    &   1.81&0.22 \\
\hline 5& CXOU J005323.8-722715&1.18    &           &  &   0.00    &   0.37    &   0.43&0.29\\
\hline 6& CXOU J005736.2-721934&0.70    &           &  &   0.47    &   0.27    &   1.14&0.39\\
\hline 7& RX J0050.8-7316& 2.78& 0.95& 33& 0.69& 0.20& 1.16&0.21 \\
\hline
\end{tabular}
\caption{Spectral information obtained using XSPEC with an absorbed powerlaw model. A distance of 60
kpc has been assumed for all objects.} \label{tab2}
\end{minipage}
\end{table*}

\subsection{CXOU J005736.2-721934}

CXOU J005736.2-721934 was originally discovered in Chandra observation 1881 on 15 May 2001 (Macomb et
al. 2003) where it was reported to have a pulse period of 565.83s. It was weakly visible in
observation 2948 where it was found to have a pulse period of $562 \pm 8.4 s$ with a confidence of
90\%.

The spectrum was obtained from data binned to 2 counts per bin.

\subsection{RX J0050.8-7316}

RX J0050.8-7316 has a well established pulse period of 323s (Coe et al, 2002). This period was also
weakly detectable in observation 2945 although at a lower power than several other peaks and would
not have been regarded as significant if the source were not already known.

\section{DISCUSSION}

\begin{figure}\begin{center}
\includegraphics[width=70mm]{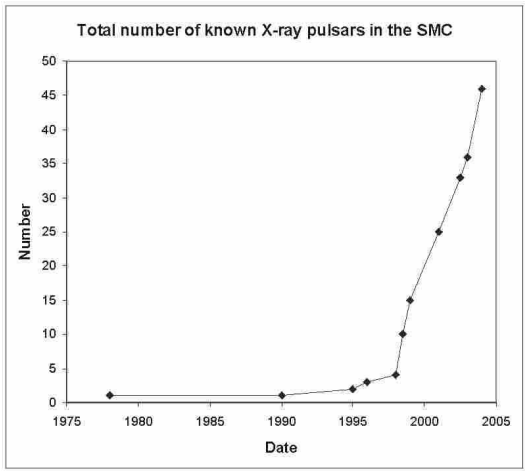}
\caption{Number of X-ray binary pulsars known in the SMC.}
\label{fig:pn}
\end{center}
\end{figure}

The discovery of these pulsars brings the total number of X-ray
binaries so far discovered in the SMC to 46. Figure~\ref {fig:pn}
shown the number of known pulsars by date and illustrates how the
steep upward trend shows no sign of levelling off. The research of
Naz{\' e} et al.(2003) points to the possibility that there may be
hundreds more whereas a relative comparison based on the
distribution of these systems in the Milky Way would lead one to
expect a population of perhaps one or two. Such a high density
must provide clues about the star formation rates in the SMC and
evidence from which a more accurate picture of its formation and
recent history can be developed.

The broad range of local $N_{H}$ values seen in this relatively
small sample may simply reflect differing local environments or
alternatively may indicate a measure of the depth of the source in
the SMC. The latter may be an important effect since the work of
Laney \& Stobie (1986) has shown that the SMC may have a depth of
20 kpc. Further research in this area should help to formulate a
more accurate three dimensional picture of the SMC.

\section*{Acknowledgments}

The authors are grateful to the The Chandra X-Ray Center (CXC),
which is operated for NASA by the Smithsonian Astrophysical
Observatory, for the use of the Chandra Observations.


\newpage

\begin{thebibliography}{99}

\newcommand{\mnras}{MNRAS}
\newcommand{\aap}{A\&A}
\newcommand{\aaps}{A\&AS}
\newcommand{\apss}{Ap\&SS}
\newcommand{\apjl}{APJ}
\newcommand{\aj}{AJ}
\newcommand{\apjs}{ApJS}
\newcommand{\iaucirc}{IAUC}
\newcommand{\pasj}{PASJ}
\newcommand{\pasp}{PASP}
\newcommand{\apj}{APJ}
\newcommand{\nyp}{Not yet published}

\bibitem[Coe et al. (1993)]{}Coe, M.~J., Everall, C., Norton, A.~J, Roche, P.,Unger, S.~J, Fabregat, J., Reglero, V., \& Grunsfeld, J.~M.,\ 1993, \mnras, 261,599.
\bibitem[Coe(2000)]{2000bpet.conf..656C} Coe, M.~J.\ 2000, ASP Conf.~Ser.~214:The Be Phenomenon in Early-Type Stars, 656
\bibitem[Coe et al.(2001)]{2001MNRAS.324..623C} Coe, M.~J., Negueruela, I.,Buckley, D.~A.~H., Haigh, N.~J., \& Laycock, S.~G.~T.\ 2001, \mnras, 324, 623
\bibitem[Coe et al.(2002)]{2002MNRAS.332..473C} Coe, M.~J., Haigh, N.~J.,Laycock, S.~G.~T., Negueruela, I., \& Kaiser, C.~R.\ 2002, \mnras,332, 473
\bibitem[Corbet et al.(2002)]{2002IAUC.7932....2C} Corbet, R., Markwardt, C.~B., Marshall, F.~E., Laycock, S., \& Coe, M.\ 2002, \iaucirc, 7932, 2
\bibitem[Corbet, Marshall, Peele, \&Takeshima(1999)]{1999ApJ...517..956C} Corbet, R.~H.~D., Marshall,F.~E., Peele, A.~G., \& Takeshima, T.\ 1999, \apj,517, 956
\bibitem[Corbet et al.(2003)]{2003HEAD...35.1730C} Corbet, R.~H.~D., Edge, W.~R.~T., Laycock, S., Coe, M.~J., Markwardt, C.~B., \& Marshall, F.~E.\ 2003, AAS/High Energy Astrophysics Division, 35,
\bibitem[]{} Corbet, R.H.D., Marshall, F.E., Coe, M.J., Laycock, S., {\&} Handler, G. 2001, AJ, 548, L41
\bibitem[Crampton, Hutchings, \& Cowley(1978)]{1978ApJ...223L..79C} Crampton, D., Hutchings, J.~B., \& Cowley, A.~P.\ 1978, \apjl, 223, L79
\bibitem[Haberl et al.(2004)]{2004ATel..219....1H} Haberl, F., Pietsch, W., Schartel, N., Rodriguez, P., \& Corbet, R.~H.~D.\ 2004, The Astronomer's Telegram, 219, 1
\bibitem[Haberl \& Pietsch(2004)]{2004A&A...414..667H} Haberl, F.~\& Pietsch, W.\ 2004, \aap, 414, 667
\bibitem[Harries, Hilditch, \& Howarth(2003)]{2003MNRAS.339..157H} Harries, T.~J., Hilditch, R.~W., \& Howarth, I.~D.\ 2003, \mnras, 339, 157
\bibitem[Laney \& Stobie(1986)]{1986MNRAS.222..449L} Laney, C.~D.~\& Stobie, R.~S.\ 1986, \mnras, 222, 449
\bibitem[Macomb, Fox, Lamb, \& Prince(2003)]{2003ApJ...584L..79M} Macomb,D.~J., Fox, D.~W., Lamb, R.~C., \& Prince, T.~A.\ 2003, \apjl,584, L79
\bibitem[Meyssonnier \& Azzopardi(1993)]{1993A&AS..102..451M} Meyssonnier, N.~\& Azzopardi, M.\ 1993, \aaps, 102, 451
\bibitem[Naz{\' e} et al.(2003)]{2003ApJ...586..983N} Naz{\' e}, Y.,Hartwell, J.~M., Stevens, I.~R., Manfroid, J., Marchenko, S.,Corcoran, M.~F., Moffat, A.~F.~J., \& Skalkowski,G.\ 2003, \apj,586, 983
\bibitem[]{} Negueruela, I.~\& Coe, M.~J.\ 2002, \aap, 385, 517
\bibitem[]{} Negueruela, I.\ 1998, \aap, 338, 505-510
\bibitem[]{}Okazaki, A.~T.\ 1993, \apss, 210, 369
\bibitem[]{} Okazaki, A.T. {\&} Negueruela, I. 2001, A{\&}A, 377, 161
\bibitem[Okazaki, Bate, Ogilvie, \& Pringle(2002)]{2002MNRAS.337..967O} Okazaki,A.~T., Bate, M.~R., Ogilvie, G.~I., \& Pringle, J.~E.\ 2002, \mnras, 337, 967
\bibitem[Porter \& Rivinius(2003)]{2003PASP..115.1153P} Porter, J.~M.~\& Rivinius, T.\ 2003, \pasp, 115, 1153
\bibitem[Rosat(2000)]{2000yCat.9030....0R} Rosat, C.\ 2000, VizieR Online Data Catalog, 9030, 0
\bibitem []{}Schwering, P.~B.~W., \& Israel, F.~P.\ 1991,A{\&}A, 246, 231
\bibitem[Tucholke, de Boer, \& Seitter(1996)]{1996A&AS..119...91T} Tucholke, H.-J., de Boer, K.~S., \& Seitter, W.~C.\ 1996, \aaps, 119, 91
\bibitem[White, Giommi, \& Angelini(2000)]{2000yCat.9031....0W} White, N.~E.,Giommi, P., \& Angelini, L.\ 2000, VizieR Online Data Catalog, 9031,
\bibitem[Yokogawa et al.(2003)]{2003PASJ...55..161Y} Yokogawa, J., Imanishi, K., Tsujimoto, M., Koyama, K., \& Nishiuchi, M.\ 2003, \pasj, 55, 161
\bibitem[]{} Zezas, A., McDowell, J.C., Hadzidimitriou, D., Kalogera, V., Fabbiano, G., \& Taylor, P., \ 2003. \ astro-ph/0310562
\bibitem[Zaritsky et al.(2002)]{2002AJ....123..855Z} Zaritsky, D., Harris, J., Thompson, I.~B., Grebel, E.~K., \& Massey, P.\ 2002, \aj, 123, 855
\end{thebibliography}
\end{document}